\documentclass[final,3p,times,twocolumn]{elsarticle}

 \biboptions{comma,sort&compress}

\usepackage{here}
 \usepackage{graphicx}
  \usepackage{epsfig}

\def\als{\alpha_{\rm s}} 
\def\MS{\overline{\rm MS}}
\def\nin{\noindent}
\def\beq{\begin{equation}}
\def\eeq{\end{equation}}
\def\bea{\begin{eqnarray}}
\def\eea{\end{eqnarray}}

\journal{Nuc. Phys. (Proc. Suppl.)}

\begin{document}

\begin{frontmatter}

\title{$\alpha_s$ from the static energy in QCD}

\author[ITP]{Xavier Garcia i Tormo}

\address[ITP]{Albert Einstein Center for Fundamental Physics. Institut f\"ur Theoretische Physik, Universit\"at Bern,
  Sidlerstrasse 5, CH-3012 Bern, Switzerland}

\begin{abstract}
\noindent
Comparing perturbative calculations with a lattice computation of the static energy in quantum chromodynamics at short distances, we
obtain a determination of the strong coupling $\alpha_s$. Our
determination is performed at a scale of around 1.5 GeV (the typical
distance scale of the lattice data) and, when evolved to the $Z$-boson mass scale $M_Z$, it corresponds to $\alpha_s\left(M_Z\right)=0.1156^{+0.0021}_{-0.0022}$.

\end{abstract}

\end{frontmatter}

This talk is based on Ref.~\cite{Bazavov:2012ka}, to which we refer
for additional details.

The energy between a static quark and a static antiquark that are
separated a distance $r$, i.e. the quantum chromodynamics (QCD) static
energy, $E_0(r)$, is a good object to study in order to understand the
behavior of the theory. One can identify a long-distance part and a
short-distance part of the static energy, both of which can be
computed with lattice simulations. Here we will focus only on the short-distance part,
i.e. on distances $r\lesssim 0.234$ fm, where weak-coupling
calculations are also reliable. The comparison of the lattice computation
with the perturbative calculation tests our ability to describe the
short-distance regime of QCD, and provides information on the
region of validity of the weak-coupling approach. This comparison also
allow us to determine the strong coupling $\alpha_s$, which is the
subject of this talk. There has been a lot of recent activity
regarding both, the lattice computations and the perturbative
calculations of the static energy, which has allowed for a precise
determination of $\alpha_s$ from it to be possible \cite{Bazavov:2012ka}.

On the perturbative side, the static energy is known, at present,
including terms up to order $\als^{4+n}\ln^n\als$ with $n\ge 0$ \cite{Brambilla:2010pp,Pineda:2011db,Brambilla:2009bi,Smirnov:2009fh,Anzai:2009tm,Smirnov:2008pn,Brambilla:2006wp}. That
is, three-loop with resummation at sub-leading accuracy of the
$\ln\als$ terms that appear at short distances. We refer to this level
of accuracy as next-to-next-to-next-to leading-logarithmic (N$^3$LL).

On the lattice side, the static energy has recently been calculated in
$2+1$ flavor QCD \cite{Bazavov:2011nk}, using a combination of tree-level improved gauge
action and highly-improved staggered quark action \cite{Follana:2006rc}. This
computation employed the physical value for the strange-quark mass $m_s$
and light quark masses equal to $m_s/20$, which correspond to a pion
mass of about 160 MeV in the continuum limit, very close to the
physical value. The computation was performed for a wide range of gauge
couplings, and was corrected for lattice artifacts. It allows to study
the static energy down to distances $r\simeq 0.065$ fm.

The perturbative expressions for the static energy depend on the value of the QCD scale $\Lambda_{\MS}$ (in the $\MS$
scheme), and we can use the comparison with lattice data to determine
it. For that, we assume that
perturbation theory (after implementing a cancellation of the leading
renormalon singularity) is enough to describe lattice data in the range of
distances we are considering. Then, the general idea is that we can search for the values of
$\Lambda_{\MS}$ for which the agreement with lattice improves when the
perturbative order of the calculation is increased; and in that way
find the values of $\Lambda_{\MS}$ that are allowed by lattice
data. This same program was already performed for the quenched case in
Ref.~\cite{Brambilla:2010pp}; the unquenched computation of the static
energy in Ref.~\cite{Bazavov:2011nk} allows us to do the same here in
the unquenched case, and therefore obtain a value for $\alpha_s$. 

The static energy on the lattice is calculated in units of the scales $r_0$ or $r_1$, defined as \cite{Sommer,milc04}
\begin{equation}
r^2 \frac{d E_0(r)}{d r}|_{r=r_0}=1.65,~~~r^2 \frac{d E_0(r)}{d r}|_{r=r_1}=1;
\end{equation}
we use the values of $r_0$ or $r_1$ in Ref.~\cite{Bazavov:2011nk} to
obtain $\Lambda_{\MS}$ in physical units. In the
perturbative calculation one needs to implement a scheme that cancels
the leading renormalon singularity \cite{Beneke:1998rk}\footnote{In
  the lattice computation the results calculated at different lattice
  spacings are normalized to a common value at a certain
  distance.}. This kind of schemes introduce an additional dimensional scale in the problem (that we
denote as $\rho$). We implement the renormalon cancellation according
to the scheme described in Ref.~\cite{Pineda:2001zq}; then, the natural
value of the scale $\rho$ is at the center of the
range for which we have lattice data. But since any value of $\rho$
around this natural value cancels the renormalon, we can exploit this
freedom to search for a set of $\rho$ values that allow for an optimal
determination of $r_0\Lambda_{\MS}$. To obtain our central value for
$r_0\Lambda_{\MS}$ we let $\rho$ vary around its natural value; then,
for each value of $\rho$ and at each order in the perturbative
expansion, we perform a fit to the lattice data ($r_0\Lambda_{\MS}$ is
the parameter of the fits); and finally select
the $\rho$ values for which the reduced $\chi^2$ of the fit decreases
when increasing the perturbative order. Our central value for
$r_0\Lambda_{\MS}$ is then given by the average (weighted by the
inverse $\chi^2$) of those fit values\footnote{Note that the absolute value of the $\chi^2$ (from the fits of the
perturbative expressions to the lattice data), at a given order in
perturbation theory, does not have any particular significance. The
reason for that is that, in principle, one does not know exactly how
large the terms at the next perturbative order will be. Therefore, one
does not know how accurately the perturbative expression should
describe the lattice data at a given order. For that reason, our
procedure to determine $r_0\Lambda_{\MS}$, described above, does not
use absolute values of the $\chi^2$s, but rather comparisons between
$\chi^2$ values at different perturbative orders. Otherwise there
would be the danger that one artificially reduces the $\chi^2$ at a low
perturbative order, by using a ``wrong'' value of $r_0\Lambda_{\MS}$
that is not suitable for a more precise expression at a higher
perturbative order.}. We can perform the above
analysis at different orders of accuracy; at N$^3$LL accuracy the
perturbative expression depends on an additional constant (due to the
structure of the renormalization group equations at this order), which
would also need to be fitted to the lattice data. When we do the fits
at N$^3$LL accuracy, we find that the $\chi^2$ as a function of $r_0\Lambda_{\MS}$ is very
flat; we interpret this fact as the data not being
sensitive to the sub-leading ultrasoft logarithms. Therefore we take
the result at three loops with resummation of the leading ultrasoft
logarithms as our best result. For illustration, we present here the
expressions for the static energy at this level of accuracy. The static energy
at next-to-next-to-next-to leading order (N$^3$LO) is given by
\bea\label{eq:E0N3LO}
&& \hspace{-8mm} 
E_0^{{\textrm{\tiny{N}}}^3{\textrm{\tiny{LO}}}}(r) =-\frac{C_F\als(1/r)}{r}
\Bigg\{1 + \tilde{a}_1\,\frac{\als(1/r)}{4\pi}
\nonumber
\\ 
&& \hspace{-6mm}  
+ \tilde{a}_{2}\,\left(\frac{\als(1/r)}{4\pi}\right)^2  
+ \Big[\frac{16\,\pi^2}{3} C_A^3 \,  \ln {\frac{C_A\als(1/r)}{2}}
\nonumber
\\ 
&& \hspace{-6mm}  
+ \tilde{a}_{3}
  \Big]\! \left(\frac{\als(1/r)}{4\pi}\right)^3\Bigg\}+K_1,
\eea
with (numerically, for $N_c=3$)
\bea
\hspace{-4mm}
\tilde{a}_1&=&23.032 - 1.8807 n_f, 
\\
\hspace{-4mm}
\tilde{a}_{2}&=& 1396.3 - 192.90 n_f + 4.9993 n_f^2,
\\
\hspace{-4mm}
\tilde{a}_{3}&=&
108654. - 21905.2 n_f + 1284.69 n_f^2 
\nonumber
\\ 
&&
- 20.6009 n_f^3,
\eea
where $C_F=(N_c^2-1)/(2N_c)$, $C_A=N_c$, $n_f$ is the number of light
flavors (i.e. $n_f=3$ in our case), and $K_1$ is a constant that, in the
comparison with data, gets absorbed in the constant used to make the static energy coincide
with the lattice point at the shortest distance available. If we include
the resummation of the leading ultrasoft logarithms we have
\bea
E_0^{{\textrm{\tiny{N}}}^3{\textrm{\tiny{LO}}}+{\textrm{\tiny{us.res.}}}}(r)
& = &\!\!\!
\Bigg\{\textrm{Eq.}~(\ref{eq:E0N3LO})+\frac{C_F\als^4(1/r)}{r}\frac{1}{12\pi}
C_A^3 \nonumber\\
&&\hspace{-25mm}\times \ln {\frac{C_A\als(1/r)}{2}}\Bigg\}+\frac{2C_FC_A^3}{12\beta_0r}\als^3(1/r)\ln\frac{\als(\mu)}{\als(1/r)}\nonumber\\
&&\hspace{-25mm}
-\frac{C_FC_A^3}{12\pi r}\als^4(1/r)\ln\frac{C_A\als(1/r)}{2 r\mu},
\eea
where $\mu$ is the ultrasoft scale (of order $\als/r$), and
$\beta_0=(11/3)C_A-(4/3)T_Fn_f$, with $T_F=1/2$. To implement the
required renormalon cancellation we use the so-called RS scheme
\cite{Pineda:2001zq}. That is, if we calculate the static energy at
$m$-loop order in perturbation theory we add the following term to it \cite{Brambilla:2009bi}
\bea  
\textrm{RS\small{subtr.}} & = & R_{s} \, \rho \,  \sum_{n=1}^m \left(
\frac{\beta_0}{2\pi} \right)^n \als(\rho)^{n+1} \nonumber\\
&&\times\sum_{k=0}^2
d_{k} \frac{\Gamma(n+1+b-k)}{\Gamma(1+b-k)}\,,
\eea
with $R_s=-1.123$ the normalization of the $u=1/2$ renormalon
singularity (which we computed approximately according to the
procedure in Ref.~\cite{Lee:1999ws}), and
\bea
d_0 & = & 1\, ,\nonumber\\
d_1 & = & \frac{\beta_1^2-\beta_2\beta_0}{4 b
   \beta_0^4}\, ,\nonumber\\
d_2 & = & \frac{-2 \beta_0^4 \beta_3+4 \beta_0^3
   \beta_1 \beta_2}{32 (b-1) b
   \beta_0^8}\nonumber \\
&&+\frac{\beta_0^2
   \left(\beta_2^2-2 \beta_1^3\right)-2 \beta_0 \beta_1^2 \beta_2+\beta_1^4}{32 (b-1) b
   \beta_0^8}\, ,
\eea
with
\begin{equation}
b=\frac{\beta_1}{2\beta_0^2}\, ,
\end{equation} 
(the higher order coefficients of the beta function, $\beta_{1,2,3}$,
can be found, for instance, in
Refs.~\cite{vanRitbergen:1997va,Czakon:2004bu}).

Having determined our central value for $r_0\Lambda_{\MS}$ we now need to
assign an error to it. The error must reflect the uncertainties
associated to the neglected higher-order terms in the perturbative
expansion. To account for that, we consider the weighted standard
deviation in the set of $\rho$ values we found before, and the
difference with the weighted average computed at the previous
perturbative order. The latter term turns out to be the dominant error; we
then add the two errors linearly. Additionally, we also redo the
analysis with alternative weight assignments ($p$-value, and constant
weights); we obtain compatible results, and quote and error that
covers the whole range spanned by the three analyses. As a further
cross-check, we can compare the analysis performed with the static energy normalized in
units of $r_0$ (our default choice) and the one with the static energy
normalized in units of $r_1$; we find that the two analyses give
consistent results\footnote{This is a cross-check and not just a
  trivial rescaling, because the errors and systematics entering the lattice
analyses normalized in units of $r_0$ or $r_1$ are different.}. 

Our final result reads
\begin{equation}\label{eq:Lambda}
r_0\Lambda_{\MS}=0.70\pm0.07,
\end{equation}
which using the value of $r_0$ from Ref.~\cite{Bazavov:2011nk} gives
\begin{equation}\label{eq:as1p5}
\alpha_s\left(\rho=1.5{\rm GeV},n_f=3\right)=0.326\pm0.019.
\end{equation}
When we evolve Eq.~(\ref{eq:as1p5}) to the $Z$-mass scale, $M_Z$, we obtain
\begin{equation}\label{eq:asMZ}
\alpha_s\left(M_Z,n_f=5\right)=0.1156^{+0.0021}_{-0.0022},
\end{equation}
where we have used the \verb|Mathematica| package \verb|RunDec| \cite{Chetyrkin:2000yt} to obtain
the above number (4 loop running, with the charm quark mass equal to
1.6 GeV and the bottom quark mass equal to 4.7 GeV).

\begin{figure}
\centering
\includegraphics[width=7.5cm]{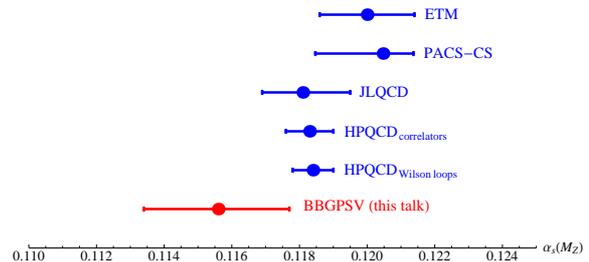}
\caption{Comparison of our result (red -lighter- point) with other
  recent lattice determinations of $\als$ (blue -darker- points). The
  references are: HPQCD \cite{McNeile:2010ji}, JLQCD
  \cite{Shintani:2010ph}, PACS-CS \cite{Aoki:2009tf}, ETM \cite{Blossier:2012ef}.
}\label{fig:compas}
\end{figure}

We compare our result with other recent lattice determinations of
$\als$ in Fig.~\ref{fig:compas}. Our central value is a bit lower than
those of the other lattice determinations.

Our determination is performed at a scale of around 1.5 GeV. This
scale corresponds to (the inverse of) the typical distance where: (i) we have
lattice data, and (ii) the weak-coupling calculation is reliable. This
means that our analysis represents the lowest-energy determination of $\als$
available, and can therefore be an important new ingredient to further
test the running of $\als$. Previously, the lowest-energy determination was that
coming from hadronic $\tau$ decays (performed at $m_{\tau}=1.78$
GeV). For comparison, the value of the pre-average of $\als$
determinations from $\tau$ decays that is currently used by the
Particle Data Group (PDG) \cite{Beringer:1900zz} is
$\als(M_Z)=0.1197\pm0.0016$. 

\begin{figure}
\centering
\includegraphics[width=7.5cm]{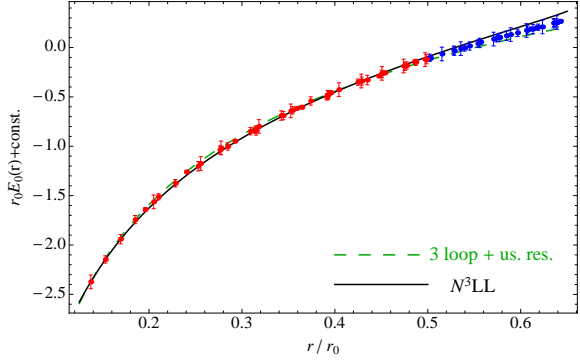}
\caption{Comparison of the singlet static energy with lattice data
  (red -lighter- points). [The comparison (and all the analysis in the text) is
  done for $r<0.5r_0\simeq0.234$ fm, which is the region where
  perturbation theory is reliable. The (blue -darker-) points and curves for $r>0.5r_0$
  are shown just for illustration]. The long-dashed green curve is at
  three loops plus leading ultrasoft logarithmic resummation, and the
  solid black curve also includes resummation of the sub-leading
  ultrasoft logarithms (i.e. it is at N$^3$LL accuracy). $r_0\Lambda_{\MS}=0.70$ was used
  in all the curves. The additive constant in the perturbative
  expression for the static energy is
  taken such that each curve coincides with the lattice data point at
  the shortest distance.
}\label{fig:complatt}
\end{figure}

To summarize, we have obtained a determination of $\als$ by comparing
perturbative calculations with a lattice computation of the
short-distance part of the QCD static energy. Our determination is at
three-loop accuracy (including resummation of the leading ultrasoft
logarithms), and is performed at a scale of 1.5 GeV (and therefore it
constitutes the lowest-energy determination of $\als$ available). When
evolved to the scale $M_Z$, it corresponds to
$\alpha_s\left(M_Z\right)=0.1156^{+0.0021}_{-0.0022}$. A very
important outcome of our work is also that our analysis shows, for the first time in QCD with $n_f=2+1$
flavors, that perturbation theory (after cancellation of the leading
renormalon singularity) can describe the short-distance part of the
static energy. This is illustrated in Fig.~\ref{fig:complatt}.

\section*{Acknowledgements}
\nin
It is a pleasure to thank Alexei Bazavov, Nora Brambilla, P\'eter
Petreczky, Joan Soto, and Antonio Vairo for collaboration on the work
reported in this talk. I also thank P. Petreczky for comments on the present
manuscript. 


\end{document}